\documentstyle[12pt,twocolumn,mmatext]{article}
\title{Generalized States for Multipion Production and Disoriented Chiral 
Condensates}
\author{B.Bambah,\\ School of Physics , Univ. of Hyderabad,\\ Hyderabad, 
500 046, India}
\date{-}
\begin{document}
\maketitle
\begin{abstract}
We show that the sudden quenching mechanism responsible for the production of 
the
disoriented chiral condensate gives rise automatically to squeezed states.
We compare the distribution of charged and neutral pions in the two extreme 
limits
of a sudden quench and a slow adiabatic relaxation and show that in the former
case the difference in the distributions is much more dramatic than in the 
latter.
We also examine the isospin stucture of the resulting squeezed pion 
distributions.
\end{abstract}

\section{Introduction}
In high energy collisions, besides the quark-hadron phase transition, the 
particular
phase transition exciting great interest these days is chiral phase 
transition .
Following the pioneering work of Bjorken \cite{bj92}, Kowalski, Taylor 
\cite{kow92}, Wilczek and Rajagopal \cite{Wil93}, there
exists the exciting possibility that in the upcoming high luminosity 
hadron and heavy ion collisions 
, after a chiral phase transition, a disoriented chiral condensate (DCC) 
may be formed.
The physics of the formation mechanism in various low energy models of 
QCD has been 
presented in the talk of Dr. A. M. Srivastava in these proceedings.
Theoretical treatments of the DCC have been 
based on coherent pion states and recently Amado and Kogan \cite{Ama88} 
have shown that 
a mean field quantization of the linear $\sigma$ model (which is one of the
canonical effective models of QCD used in the description of the DCC) leads 
to a
squeezed state description of the DCC at the quantum level. In this paper
we examine this description in more detail and show that the amount of 
squeezing 
which leads to an amplification of the low-momentum pion modes is directly 
related
to the mechanism by which the system goes through the phase transition.
We show that if the transition is a sudden quench the squeezing and 
amplification is much more dramatic
than if the transition is a slow adiabatic one.

\section{Quenching and Squeezing}
 In the quantum treatment of sigma model in mean field approximation
the mechanism for amplification of low momentum modes is done by 
considering the
equation of motion for the pion field
with $\phi^a\phi_a=\langle\phi_{a}\phi^{a}\rangle(t)$.
\begin{equation}
\frac{\partial^2 \pi(\overline{k},t)}{dt^2}+
[k^2+\lambda(\langle\phi^2\rangle(t)-v^2)]\pi(\overline{k},t)=0
\end{equation}
For initial condition $\langle \phi^2 \rangle <v^2$ long wavelength modes 
grow exponentially and $\langle \phi \rangle$
oscillates around  $v^2$ damps down to stability.
Each long wavelength mode becomes parametrically excited 
oscillator  and amplification of zero point oscillations
gives DCC. Hence coherent state descriptions are  often used.

In Mean Field approximation with pion wave  function $|\Psi>=\Pi_{i,k}
|\psi>_{i,k} $
the equation of motion for the pion field can be viewed as a time dependent 
oscillator equation
with a time dependent frequency $\Omega(k,t)=\overline{k}^2+\lambda(<\phi^2>
(t)-v^2)$ evolving by the Hamiltonian
\begin{equation}
H_{k}(t)=[\frac{1}{2}(P_{\pi}^2+\Omega^2(k,t)\pi^2(k))]
\end{equation}
 Creation
and annihilation operators of the observed pions in the decay of DCC are 
given by
$a(k)=\frac{w(k)\pi(k)+iP_{\pi}}{\sqrt{2w(k)}}$ and 
$a^{\dag}(k)=\frac{w(k)\pi(k)-iP_{\pi}}{\sqrt{2w(k)}}$ 
with $w(k)=\Omega(k,\infty)=\sqrt{k^{2}+m_{\pi}^{2}}$.
The Hamiltonian $H_{k}(t)$ can be written as: 
\begin{equation}
H_{k}(t)=\Omega_{k,t} A^{\dag}(k,t)A(k,t)
\end{equation}
Where $A,A^{\dag}$ are related to $ a, a^{\dag}$ by the Unitary evolution 
operator corresponding to
$H_{k}(t)$ i.e. \cite{lo}
\begin{equation}
A(k,t)=U(t,t_0)a(k)U^{\dag}(t,t_{0})
\end{equation}
it can be shown using SU(1,1) algebra that this reduces to a Bogolubov 
(squeezing) transformation
$A=\mu(t) a(k)+\nu(t) a^{\dag}(k)$
with $\mu=Sinh(r)$ and $\nu=Cosh(r)$, where $r$ is the squeezing parameter 
and $\mu$ and $\nu$ are functions of the sums and differences of
$w(k)$ and $\Omega(k,t)$.
In the case of the DCC an initial vacuum $|0>$ is transformed to :
\begin{equation}
S(z)=e^{\frac{1}{2}(z^*a^2-za^{\dag 2})}|0>
\end{equation}
$z=re^{i\theta}$  \cite{yue76}.

The wave function $|\psi_{k}>$ evolves as
\begin{equation}
i\frac{\partial|\psi>_{k}}{\partial t}=H_{k}(t)|\psi>_{k}.
\end{equation}
In the co-ordinate representation the equation of motion for $\pi$ can
be viewed as a Shrodinger Equation for $\psi$ with a time dependent potential
$V(t)=-\lambda(<\phi^2>(t)-v^2)$
so
\begin{equation}
-\psi''+V(t)\psi=\omega^2(k)\psi
\end{equation}
For a general time dependent oscillator with variable frequency where V(t) 
takes
constant values at $t\longrightarrow \pm \infty$ the problem becomes one of 
a potential
barrier reflection and it can be shown that the transmission coefficient of 
a wave passing through the barrier
can be related to the squeezing parameter in a particle creation problem by 
the relation
\begin{equation}
Sinh^2r=\frac{T-1}{T}=<n>
\end{equation}
Now we are in a position to examine the problem of quenching versus 
adiabaticity for squeezing.
Let us take $<\phi^2>(t)$
given by:
\begin{eqnarray}
<\phi >(t)= \nonumber \\
f_{\pi}[Tanh[(t-\tau)/a]-Tanh[-\tau/a]]
\end{eqnarray}
Then the limit $a->0$ corresponds to the quenching limit  and $a->\infty$ 
corresponds to the adiabatic limit  as shown in fig.1.
Here $\tau$ represents the damping time of  $<\phi(t)>$ and $a$ is given 
in units of $\tau$. Typically $\tau$ is taken to be
$3-6 m_{\sigma}^{-1}$.
The corresponding potential barrier is of the modified
 Poschl Teller Potential given in fig.2 which
in the quenched limit  corresponds to the rectangular barrier.
The potentials in both limits are exactly soluble and the value of the 
transmission coefficients as a function of k
can be calculated . Using these  and equation 8
fig. 3 shows the variation of $<n>$ with k for various values of a.
In the quenching limit, the long wavelength modes are clearly amplified 
and the squeezing is very pronounced. In the
adiabatic limit no substantial squeezing is present. This clearly illustrates 
the connection between quenching and squeezing.
A detailed calculation tabulating various forms of $<\phi(t)>$ and the 
resulting amplification will be
published in a later communication.

An alternate way to vary the frequency of the time dependent oscillator would 
be to take the pion mass as an exponential function of temperature
and examine the evolution in terms of thermal state in terms of a time 
dependent temperature. The squeezed state
formalism is particulary useful for this and the methods of thermofield 
dynamics (TFD) can be used effectively. This work
is in progress.
\begin{figure}
\fbox{\input{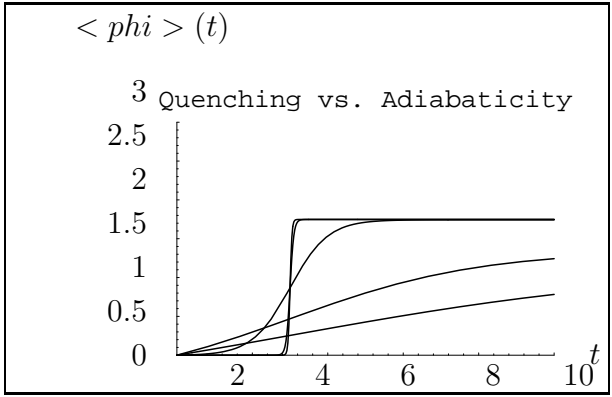}}
\caption{ Shows the variation of $<\phi(t)>$ with t for various values of a.}
\end{figure}
\begin{figure}
\fbox{\input{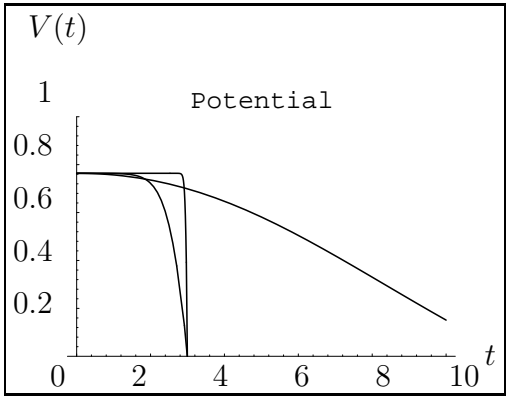}}
\caption{ Shows potential V(t) for a=$10^{-4}$(rectangular barrier),.1
(gradual quench) ,1 (Poschl-Teller).}
\end{figure}
\begin{figure}
\fbox{\input{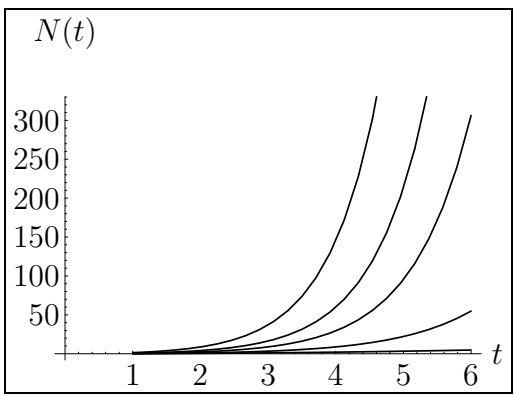}}
\caption{ Shows the variation of $<n>$ with wave number k for values of 
a =.0001 , .1,1 ,10. ,.}
\end{figure}
\section{Pion Radiation from DCC}
Having shown that the mean Field treatment of DCC suggests Squeezed State 
Treatment of emerging pion waves and that this effect
is more pronounced in quenching we now proceed to construct most general 
class of ispospin squeezed states which can produce the phenomenology
of DCC decay based on first principles.
The most general eigenstate of $I$ and $I_3$ and  having
properties of  squeezing can be constructed by
\begin{equation}
F(A,A^{\dag},N)Y^{l}_{m}(a^{\dag})|0>.
\end{equation}
where
$F=e^{\frac{1}{2}(fA^{\dag}-f^{*}A)} $.
\begin{equation}
|\psi,f>=e^{\frac{1}{2}(fA^{\dag}-f^{*}A)} Y^{l}_{m}(a^{\dag})|0>
\end{equation}
The state $|\psi,f>$ is the correct  displacement operator
coherent state of definite isospin \cite{BS},\cite{HS71}. 
Where
 bilinear operators $A,A^{\dag}$ and $N$ which commute  
with  $I$ and $I_3$, 
in terms of pion creation and ahnnihilation operators are:
\begin{eqnarray}
A=\vec{a}\cdot\vec{a}=a_{0}a_{0}-2a_{+}a_{-} \nonumber \\
A^{\dag}=\vec{a^{\dag}}\cdot \vec{a^{\dag}}=a_{0}^{\dag}
a_{0}^{\dag}-2a_{+}^{\dag}a_{-}^{\dag} \nonumber \\
N=a_{0}^{\dag}a_{0}+a_{1}^{\dag}a_{1}+a_{2}^{\dag}a_{2} 
\end{eqnarray}
The multiplicity distributions for  $\pi_{0}$,$\pi^{+}$ and $\pi^{-}$
can be written as:
\begin{eqnarray}
|<n_{0},n_{+},n_{-}|\psi,f,l,m>|^{2}=\nonumber \\
(-\frac{1}{2})^{2m}[(2l+1)(l-m)!(l+m)!] \nonumber \\
|\sum_{k=0}^{(l-m)/2}\frac{2^{\frac{m}{2}-k}}{(l-m-2k)! k!(m+k)!}\nonumber \\
S_{n_{0},l-m-2k}S^{Two-mode}_{n_{+},n_{-},m+k,k}|^{2}
\end{eqnarray}

where $S(f)$ is the one mode squeezing operator
\begin{eqnarray}
S(f)=&\nonumber\\
 <n_{0}|e^{\frac{1}{2}(f(a_{0}^{\dag})^{2}-f^{*}a_{0}^{2})}|l-m-2k>\nonumber\\
= S_{n_{0},l-m-2k}. 
\end{eqnarray}
$S^{t}(f)$ is then the two mode squeezing operator
\begin{eqnarray}
<n_{+},n_{-}|e^{(fa_{+}^{\dag}a_{-}^{\dag}-f^{*}a_{+}a_{-})}|m+k,k>\nonumber\\
=S_{n_{+},n_{-},m+k,k}. 
\end{eqnarray}

For $p\overline{p}$ collisions, isospin conservation requires $I=0,1$ states. 
For
 $pp$ collisions
$I=0,1,2 $ \\
 $\pi p$ collisions can have upto $I=3$.
  Charge conservation implies that
$n_{\pi^{+}}=n_{\pi^{-}}$ 
so that we have predominantly $I_{3}=0$ states.\\
For $I=0, I_3=0$ ,$n_{+}=n_{-}$, the distribution reduces to 
 the product of squeezed distributions for charged and neutral pions and only 
even number
of pions emerge.
The distribution of charged particles is 
\begin{equation}
P_{n_{c}}{}=  \frac{(tanh(f))^{2n_{c}}}{(cosh(f))^{2}}
\end{equation}
and for neutral particles:
\begin{equation}
P_{n_{0}}=\frac{n_{0}!(tanh(f))^{n_{0}}}{((\frac{n_{0}}{2})!)^{2}cosh(f) 
2^{n_{0}}}
\end{equation}
and corresponds to the product of distributions of a one-mode and two-mode 
squeezed state.
Case $I=1;m=0$: Again, $n_{+}=n_{-}=n_{c}$.
\begin{eqnarray}
|<n_{0},n_{c}|\psi,f,1,0>|^{2}=\nonumber \\
|S_{n_{0},1} S^{Two-mode}_{n_{+},n_{-},0,0}|^{2} 
\end{eqnarray}
so the neutral pion distribution is affected by isospin and not the charged 
pion distribution.
$n_{0}$ must be odd=2m+1.
\begin{eqnarray}
|2m+1,2n_{c}|\psi,f,1,0>|^2= \nonumber\\
\frac{(2m+1)!}{(\frac{(m)!}{2})^2}  
\frac{(tan(f))^{2m+2n_c})}{cosh^5(f)} 
\end{eqnarray}

The generalized squeezed Isospin eigenstate leads 
to products of two types of squeezed states of pions , the neutral pions 
being in a
one mode squeezed state and the charged pions being in an SU(1,1) coherent or 
two-mode squeezed
state. Thus the neutral and charged pion distributions are significantly 
different as the
two types of states have different properties. We now illustrate the effect 
of quenching versus adiabaticity
on these two distributions. Figs. 4 and 5 show the difference in 
the charged and neutral pion distribution for
 the adiabatic limit 
where the difference is negligible and the quenched limit where the difference 
is pronounced.
\begin{figure}
\fbox{\input{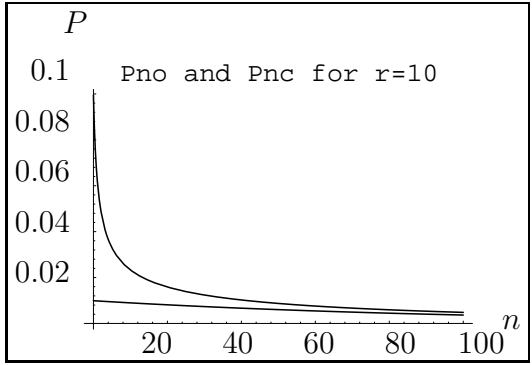}}
\caption{Shows the variation of $P_{n}{}_0$ and $P_{n}{}_c$ with $n$ for the 
adiabatic limit (r=10)}
\end{figure}
In conclusion , we have shown that the sudden quench approximation in the 
evolution
of the disoriented chiral condensate leads to a substantial amount of 
squeezing which manifests itself in the dramatic
difference between charged and neutral distributions. For an adiabatic 
quench the difference is much less, so that 
the characteristic signal examined in literature for the DCC is related 
directly to the way in which the DCC relaxes
to the vacuum in a chiral phase transition.
\begin{figure}
\fbox{\input{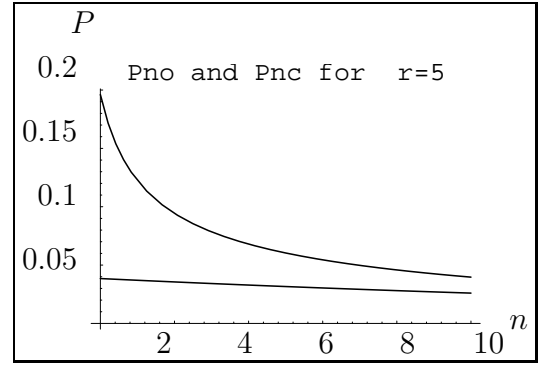}}
\caption{Shows the variation of $P_{n}{}_0$ and $P_{n}{}_c$ with $n$ for 
the quenched limit (r=5)}
\end{figure}

\end{document}